\documentclass[12pt]{article}
\usepackage[toc,page]{appendix}
\usepackage{breqn}
\usepackage{graphicx}
\usepackage{epstopdf}
\usepackage{amsmath}
\usepackage{amssymb}
\usepackage{amsfonts}
\usepackage{amsthm}
\usepackage[usenames]{color}
\usepackage{array}
\usepackage[percent]{overpic}

\usepackage[
colorlinks={},
linkcolor=black,
urlcolor=blue,
filecolor=blue,
citecolor=blue,
pdfstartview=FitV,
pdftitle={},
pdfauthor={},
pdfsubject={},
pdfkeywords={},
pdfpagemode=None,
bookmarksopen=true
]{hyperref}

\usepackage{epsfig}

\usepackage{hyperref}

\newcommand{\bea}{\begin{eqnarray}}
\newcommand{\eea}{\end{eqnarray}}
\newcommand{\ba}{\begin{array}}
	\newcommand{\ea}{\end{array}}
\newcommand{\ee}{\end{equation}}

\numberwithin{equation}{section}

\begin{document}

\begin{centering}
\thispagestyle{empty}
	\vspace{2cm}
	
	\textbf{\Large{
			 Complexity and  Multi-boundary Wormholes \vspace{.3cm}\\
		 in $ 2+1 $ dimensions
		 }}
	
	\vspace{0.8cm}
	
	{\large Hamed Zolfi }
	
	\vspace{0.7cm}
	
	\begin{minipage}{.9\textwidth}\small
		\begin{center}
			
			{\it 
School of Particles and Accelerators,
					Institute for Research in Fundamental Sciences (IPM)\\
				P.O. Box 19395-5531, Tehran, Iran	}

		\vspace{0.5cm}
		{\tt  \  hamedzolphy@ipm.ir }
		\\ 
			
		\end{center}
	\end{minipage}
	\vspace{0.8cm}
	\begin{abstract}
Three dimensional wormholes are global solutions of Einstein-Hilbert action. These space-times which are  quotients of a part of global AdS$ _{3} $ have multiple asymptotic regions, each with conformal boundary $ S^{1}\times\mathbb{R} $, and separated from each other by horizons.	Each outer region is isometric to BTZ black hole, and behind the horizons, there is a complicated topology. The main virtue of   these geometries is that they  are dual to known CFT states.
In this paper, we evaluate the full time dependence of holographic complexity for  the  simplest case of $ 2+1 $ dimensional Lorentzian
wormhole spacetime, which has three asymptotic AdS boundaries, using the “complexity equals
volume" (CV) conjecture. We conclude that the growth
of complexity is non-linear and saturates at late times.

	\end{abstract}
\end{centering}
\newpage
\tableofcontents
\setcounter{equation}{0}
\setcounter{page}{1}
\section{Introduction}

In recent years, some in-depth connections have been discovered between
quantum information theory and quantum gravity.
The AdS/CFT duality provides a fruitful framework for studying these connections. The prime example of such a relationship is the Ryu-Takayanagi
formula which provides a geometrical interpretation for  entanglement entropy in a dual CFT \cite{Ryu:2006bv}. Van
Raamsdonk also strengthened this relationship \cite{VanRaamsdonk:2010pw}. He argued that the amount of entanglement between two regions is related to their distance and we can connect geometries by entangling  degrees of freedom and separate them by disentangling. Later, this observation led to ER=EPR conjecture \cite{Maldacena:2013xja}. 
The next example comes from the reconstruction of a bulk
operator as a set of non-locally smeared CFT operators \cite{Hamilton:2005ju,Hamilton:2006az,Hamilton:2006fh} which leads to several paradoxes.  To resolve
these paradoxes, the authors of \cite{Almheiri:2014lwa} used the concept of
quantum error correcting code. 
 The third connection between quantum gravity and
quantum information theory is quantum computational complexity \cite{Susskind:2014moa}. 
These ideas emerged from a puzzle to understand  the growth of the Einstein-Rosen bridge for AdS black holes  in  thermal equilibrium.
Holographic
complexity   equips us to understand the rich geometric structures that exist behind the horizon.  Since a characteristic property of
quantum complexity is that it continues to grow even long times after the
boundary theory reached thermal equilibrium.
 In fact, the complexity is conjectured to continue growing until a time
scale that is exponential in the number of degrees of freedom in the system \cite{Susskind:2014rva,Susskind:2014jwa,Stanford:2014jda}. 
Quantum computational complexity is a notion from quantum information theory which
estimates the difficulty of constructing a desired target state from simple elementary gates.
In this notion, the gates are unitary operators which can be taken
from a universal set \cite{Arora:2009, Moore: 2011}.

In the context of the AdS/CFT correspondence, two proposals have been made to
evaluate the complexity of a boundary state. The first one is that the complexity should
be dual to the volume of the extremal codimension-one bulk hypersurface $ \Sigma $ which meets
the asymptotic boundary on the time slice where the boundary state is defined. This statement summarizes to: 
\begin{equation}
\mathcal{C}_{\text{V}}=\max\left[  \frac{\mathcal{V}_{\Sigma}}{G\ell}\right],
\end{equation}
where $ \ell $ is a certain length scale associated with the geometry, usually selected to be
the AdS radius of curvature or Schwarzschild radius.\footnote{In the following we will set $ G=\ell=1 $.}
According to the “complexity=action” proposal (CA), the quantum computational complexity of a
holographic state is given by the on-shell action evaluated on a bulk region known as the ‘‘Wheeler De Witt’’ (WDW) patch \cite{Brown:2015bva,Brown:2015lvg},
\begin{equation}
	\mathcal{C}_{\text{A}}= \frac{I_{\text{WDW}}}{\pi\hbar }. 
\end{equation}
Here the WDW patch is defined as the domain of dependence of any Cauchy surface in the bulk
whose intersection with the asymptotic boundary is the time slice $ \Sigma $. 

An important feature of quantum complexity is that it grows  with time. This growth  is linear with the
slope given by Lloyd’s bound \cite{Lloyd}, which is twice of the energy of the state.
The linear growth of quantum complexity  at late times is such an essential property of holographic complexity that the authors of \cite{Belin:2021bga,Belin:2022xmt} a few years after CV and CA prescriptions, proposed that  every geometrical object in asymptotically Anti-de Sitter space-time  which has linear growth (and also  reproduce the switch-back effect in shock wave geometries) is  gravitational dual of complexity.\footnote{ It should be mentioned that CA conjecture in some cases does not respect Lloyd's bound, and complexity approaches a constant at  late times. In \cite{Brown:2018bms} the authors resolve the undesired late time behavior by adding an extra boundary term to the model. This in
	fact could be naturally accommodated if one considers the model as a dimensional reduction of $ 3+1 $
	dimensional Reissner-Nordstrom black hole. Another resolution for this discrepancy appeared in \cite{Akhavan:2018wla}. Motivated by $ T\bar{T} $ deformation of a conformal boundary it has been proposed  in order to have a late time behavior consistent with Lloyd's bound one is forced to have a cut off behind the horizon whose value is fixed by the boundary cut off.
	The extension of this analysis for the charged black holes and Gauss-Bonnet-Maxwell theory can be find in\cite{Hashemi:2019xeq}.}

A leading arena for the investigation of holographic complexity is the two-sided eternal BTZ black hole
 which  is also  our motivation for the present paper. This geometry which is constructed by two entangled black hole, 
is dual to the thermofield double state \cite{Maldacena:2001kr}.\footnote{ Holographic complexity
	for two-sided black holes has been calculated in \cite{Carmi:2017jqz}.} To make this more explicit, it is written in an energy eigenbasis:
\begin{equation}\label{thermo}
|\text{TFD}\left(t_L,t_R \right)\rangle =\sum_{n}e^{-\beta E_n/2}e^{-i E_n \left( t_L+t_R\right) }|E_n\rangle_{L}|E_n\rangle_{R},
\end{equation}
where $ L $ and $ R $ denote the quantum states and times associated with the left and right
boundaries, respectively.
Thermofield double state is an entangled state of two copies of the boundary CFT
and its entanglement is responsible for the geometric connection in the bulk, i.e., the
Einstein-Rosen bridge \cite{Maldacena:2013xja,Hartman:2013qma}.

One can easily visualize the generalization of the idea of Einstein-Rosen bridge by adding  genera and boundaries. The general form of  such  new and interesting objects, which  are known as Lorentzian wormholes, is shown in Figure \ref{fig1}.
In particular case, Lorentzian wormholes are  global solutions of $ 2+1 $ dimensional Einstein-Hilbert gravity:
\begin{equation}
	S=\frac{1}{16\pi G}\int d^{3}x\sqrt{-g}(R-2\Lambda),
\end{equation}
 with a negative cosmological constant $ \Lambda $.
 Such spacetimes can be
constructed as quotients of a subregion of $ \text{AdS}_{3} $ \cite{Aminneborg:1997pz,Brill:1998pr,Skenderis:2009ju}.\footnote{See also \cite{Aminneborg:1998si} for the rotating
	case.} As  explained in \cite{Skenderis:2009ju}, these geometries are associated
with a Euclidean path integral on a certain Riemann surface $ \Sigma $ which provides a natural
candidate for the dual CFT state $ |\Sigma \rangle $. With $ b $ boundaries, the state lives in the Hilbert
space $ \mathcal{H}_{1}\otimes \mathcal{H}_{2}\otimes 
...\mathcal{H}_{b}$, where $ \mathcal{H}_{a} $ is the Hilbert space of a single CFT on the cylinder. Having a basis of
energy eigenstates $ |i \rangle_{a} $  on each boundary, it can be written in general form as:
\begin{equation}\label{state}
	|\Sigma \rangle =\sum_{i,j,k,...,l}A_{i,j,k,...,l}|i\rangle_{1}|j\rangle_{2}|k\rangle_{3}...|l\rangle_{b},
\end{equation} 
where coefficients $A_{i,j,k,...,l}$ 
can be determined by a path integral over some pair of pants geometries, with the specified states on
each boundary\cite{Balasubramanian:2014hda}.

The aim of this note is to
analyze the complexity of the state $ |\Sigma \rangle $ using  CV prescription.
 Interestingly, it was found that the complexity of such states does not follow the common future of complexity and saturates at  late times. 

The remainder of this paper is organized as follows. In Section \ref{2} we will review some preliminary statements about Lorentzian wormholes in $ 2+1 $ dimensions. Then in Section \ref{3} we will use CV conjecture to compute the complexity of a state   on three boundaries of three entangled black holes connected by Einstein-Rosen bridges  as provided in Figure \ref{pants}.  
Lastly, we
offer our conclusions in Section \ref{4}.
\section{Coordinate systems for Lorentzian wormholes}\label{2}
 Lorentzian wormholes in $ 2+1 $ dimensions can be thought of as generalized eternal BTZ black holes. 
The spatial slices of an eternal BTZ black hole have a cylindrical topology, whereas, in the wormholes, the spatial slices are general two-dimensional Riemann surfaces with boundaries.
\begin{figure}[h] 
	\centering
	\includegraphics[width=1\textwidth]{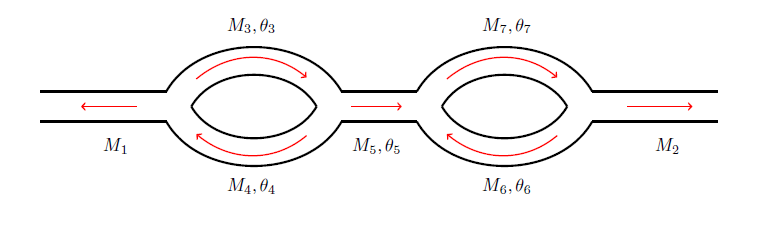}
	\caption{A fat-graph indicating a wormhole spacetime with two outer regions corresponding
		to a Riemann surface of genus 2
		with 2 boundary components.}
	\label{fig1}
\end{figure}
 For a wormhole that is based on a Riemann surface of genus $ g $ with $ b $
boundaries, one can associate a fat-graph, as indicated in Figure \ref{fig1}. A fat-graph  with $ b $ exterior  and $ (3g-3+b) $ interior edges can summarize all information about a multi-boundary wormhole. For
each outer region, there is one parameter $ M_k $, and for each inner region, two parameters
$ M_i, \theta_i $, where $  k=1, . . . ,b  $ and $ i=b+1, . . . , 3g-3+2b $. This gives a total of $ (6g-6+3b) $
parameters, showing the  number of moduli for a Riemann surface \cite{Skenderis:2009ju}. 

The spacetime is described by  two different metrics, one for the
inner charts and one for the outer charts.
The metric in the $ i $th inner patch is given by:
\begin{align}\label{inn}
	ds^2_{i}=\frac{1}{\text{cosh}^2(u) }\hspace{-.1cm}\left(\hspace{-.1cm}-du ^2\hspace{-.1cm}+\hspace{-.1cm}\frac{\alpha^{2}_i dz ^2}{(\alpha_i z+\beta_i)^{2}+\text{cos}^2\left(\theta_i \right)}\hspace{-.1cm}+\hspace{-.1cm}M_i(1+(\alpha_i z+\beta_i)^{2})d\psi^2 \hspace{-.1cm}-\hspace{-.1cm}\frac{ 2\alpha_i \sqrt{M_i}\text{sin}(\theta_i)d\psi dz}{\sqrt{({\alpha}_i z+\beta_i)^{2}+\text{cos}^2\left(\theta_i \right)}}\right),
	\end{align}
with coordinate ranges,  $ z \in[-1,1],u\in\mathbb{R}$ and  $ \psi \sim \psi + 2\pi $.
The parameters $ \alpha_i $, $ \beta_i $  are functions of the $ M_{i} $ parameters.
The metric in the $ k $th outer patch has the form:
\begin{equation}\label{out}
d s^2_{k}=\frac{\rho ^2+M_k}{\text{cosh}^2\left(\sqrt{M_k} \tau \right)} \left(-d \tau ^2+d \phi ^2\right)
+\frac{d \rho ^2}{\rho ^2+M_k}.
\end{equation}
The corresponding $ (\tau , \rho, \phi) $ coordinate system has coordinate ranges

	\begin{equation}\label{renge}
\tau\in\mathbb{R},\qquad\phi\sim\phi+2\pi,\qquad \frac{\rho  \cosh \left(\sqrt{M_k} \tau\right)}{\sqrt{M_k+\rho
		^2}}>-\frac{\beta ^2}{\beta ^2+1},
	\end{equation}
and $ \beta $ can be written in terms of $ M_k $ parameters. 
The future and past
horizons are located at:
	\begin{equation}\label{hor}
 \rho_{h}= \sqrt{M_k}\hspace{.1cm}\bigg|\sinh\left(\sqrt{M_k} \tau\right)\bigg|.
\end{equation}
The metric in the region outside of these horizons  can be written in BTZ form: 
	\begin{equation}\label{btz}
ds^2=-(r^2-M_{k})dt^2+\frac{dr^2}{r^2-M_{k}}+r^2d\phi^2,
\end{equation}
 by the coordinate transformation:
 \begin{equation}\label{bt1}
 	\tanh(\sqrt{M_k}t)=\sqrt{1+\frac{M_k}{\rho^2}}\tanh(\sqrt{M_k}\tau),
 \end{equation}
and 
	\begin{equation}\label{bt}
r^2=\frac{\rho^2+M_k}{\cosh^{2}(\sqrt{M_{k}}\tau)}.
\end{equation}
\newpage
Because it is not possible to cover the whole of the multi-boundary geometries with a single coordinate, one has to use different coordinates and metrics which have overlap in some regions. In order to have a complete prescription for describing these geometries, we need transition functions between two outers, two inners, and outer and inner charts. In the following,   transition functions between two exterior charts are presented.\footnote{Transition functions between two inner charts and also outer and inner chart can be find in \cite{Skenderis:2009ju}. }
 \subsection{Transition functions between two exterior patches}
To introduce transition functions between two exterior patches, it is useful to consider  a single pair of pants geometry constructed through the intersection of three exterior domains. These exterior
patches overlap with each other, but it can be shown
that the entire pair of pants geometry is covered by these three outer patches.
We  proceed to define transition functions on the overlapping part of the two different charts as follows:
\begin{equation}
	\tilde{\rho}=-\sqrt{\frac{M_{2}}{M_{1}}}\left( \cosh(A)\rho+\sinh(A)\cosh\left( \sqrt{M_{1}}(\phi-h)\right)\frac{\sqrt{\rho^{2}+M_{1}}}{\cosh\left(\sqrt{M_{1}}\tau \right) } \right),\nonumber
\end{equation}
\begin{equation}\label{rho}
	\sqrt{M_{1}}\tanh\left(\sqrt{M_{2}}\tilde{\tau} \right) \sqrt{\tilde{\rho}^{2}+M_{2}}=\sqrt{M_{2}}\tanh\left(\sqrt{M_{1}}\tau \right) \sqrt{\rho^{2}+M_{1}},
\end{equation}
\begin{equation}
	\exp\left(2 \sqrt{M_{2}}(\tilde{\phi}-\tilde{h})\right)=\frac{\rho\cosh(\sqrt{M_{1}}\tau)+\sqrt{\rho^{2}+M_{1}}\cosh\left( \sqrt{M_{1}}(\phi-h)-B\right)}{\rho\cosh(\sqrt{M_{1}}\tau)+\sqrt{\rho^{2}+M_{1}}\cosh\left( \sqrt{M_{1}}(\phi-h)+B\right)}.\nonumber
\end{equation}\\
Here $ M_{1} $ and $ M_{2} $ denote mass parameters in the metric on the first and the 
second patch, respectively, in  which  the transition functions were defined between them. The $M_{3} $ denotes the mass parameter
from the metric of the third patch.
Discrete parameters 
\begin{equation}\label{f}
	h=
	\left\{
	\begin{array}{ll}
		0  & \mbox{if } 1 \rightarrow 2 \\
		\pi & \mbox{if } 2 \rightarrow 1
	\end{array}
	\right.
\qquad \text{and}\qquad \tilde{h}=
	\left\{
	\begin{array}{ll}
		\pi	  & \mbox{if } 1 \rightarrow 2 \\
		0	 & \mbox{if } 2 \rightarrow 1
	\end{array}
	\right.,
\end{equation}
are responsible for avoiding left and right ambiguity on the orientation of coordinates $ \phi $ and $ \tilde{\phi} $. 
Values of $ A $ and $ B $ can be determined by: 
\begin{equation}\label{a}
	\cosh(A)=\frac{\cosh(\pi\sqrt{M_{1}})\cosh(\pi\sqrt{M_{2}})+\cosh(\pi\sqrt{M_{3}})}{\sinh(\pi\sqrt{M_{1}})\sinh(\pi\sqrt{M_{2}})},
\end{equation}
and
\begin{equation}\label{g}
	\sinh(A)\sinh(B)=1.
\end{equation}
\section{Complexity Equals Volume Conjecture}\label{3}
\begin{figure}[h]
	\centering
	\includegraphics[width=0.5\textwidth]{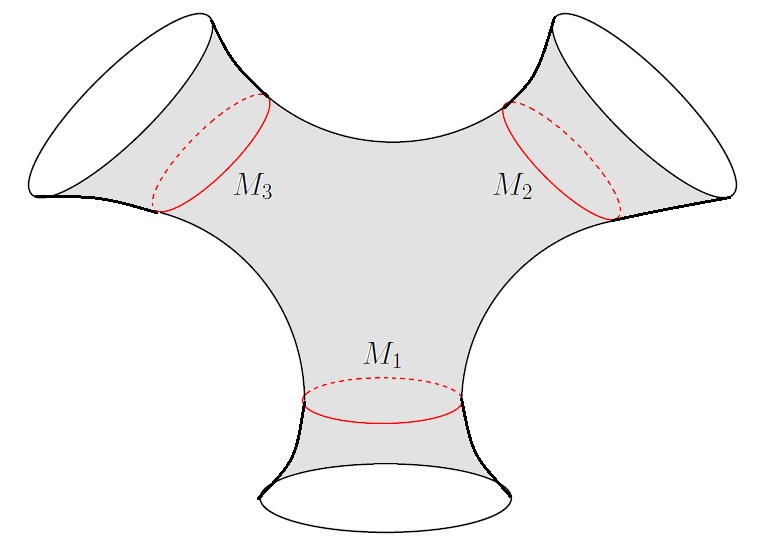} 
	\caption{ A simple case of  Lorentzian wormhole,  with $  g=0 $ and $ b=3$, represents three entangled black holes  connected by Einstein-Rosen bridges. This is reminiscent of the time slice of two-sided eternal AdS black hole, or two entangled black holes connected by a cylindrical Einstein-Rosen bridge.}
	\label{pants}
\end{figure}

All ingredients to probe  the CV  conjecture for  $\text{ AdS}_{3}$
wormhole space-times are provided.
Let us consider a simple case of generalization of thermofied double state by adding one extra asymptotic boundary region leading to a multiple
black holes geometry which are entangled in the Greenberger-Horne-Zeilinger pattern, as indicated in Figure~\ref{pants}.
\subsection{Complexity of  wormholes with three asymptotic AdS boundaries}
As  mentioned in the introduction, an important feature of these space-times is that they are dual to known CFT
states, at least in certain regions of moduli space, constructed as a path integral over
a Riemann surface with boundary. Therefore, from equation \eqref{state} the dual state of Lorentzian wormholes with three asymptotic AdS boundaries,  has the following form:
\begin{equation}\label{tripartite}
	|\Psi \rangle =\sum_{i,j,k}A_{i,j,k}|i\rangle_{1}|j\rangle_{2}|k\rangle_{3}.
\end{equation}

 In order to study the complexity of this state, one must
 evaluate the volume of the extremal codimension-one surface, whose boundaries
 correspond to the desired time slices in the three asymptotic boundaries \cite{Susskind:2014rva, Stanford:2014jda}.  Lorentzian wormhole with $  g=0 $ and $ b=3 $ does not have any inner charts, hence  only 
 the metrics of the exterior pieces should be considered. For simplicity, if  one set $ M_1=M_2=M_3=M $ in the   metric \eqref{out},  
 the induced metric of the hypersurface $ \Sigma_{k} $ for each outer  pieces will become:
	\begin{equation}\label{fact}
	d s^2_{\Sigma_{k}}=\frac{\rho ^2+M}{\text{cosh}^2\left(\sqrt{M}  \tau_{k} (\rho ) \right)} \left(-\tau_{k} '(\rho )^2 	+\frac{\text{cosh}^2\left(\sqrt{M}\tau_{k} (\rho ) \right)}{(\rho ^2+M)^{2}}\right) d \rho ^2+\frac{\rho ^2+M}{\text{cosh}^2\left(\sqrt{M} \tau_{k} (\rho ) \right)} d \phi ^2,
\end{equation}
where the primes indicate derivatives with respect to $ \rho $. Hence, the volume of Cauchy slice $ \Sigma_{k} $ which is anchored on the boundary at time $ \tau_{b_{k}} $ is:
\begin{equation}\label{volsigma}
\mathcal{V}_{\Sigma_{k}}=\int_{0}^{2\pi}\hspace{-.2cm}\int_{\rho_{\text{min}}}^{\Lambda} \sqrt{\text{sech}^2\left(\sqrt{M} \tau_{k} (\rho )\right)-\left(M+\rho
	^2\right)^2 \tau_{k} '(\rho )^2 \text{sech}^4\left(\sqrt{M} \tau_{k} (\rho
	)\right)} \,~d\rho~d\phi,
\end{equation}
where $ \Lambda $ is a boundary UV cut off.
	The complexity of the corresponding state which is located on the boundary time slice $ (\tau_{b_{1}},\tau_{b_{2}},\tau_{b_{3}}) $
  is  obtained by extremizing the volume of 
 $ \Sigma_{\text{total}}=\Sigma_{1}\cup\Sigma_{2}\cup\Sigma_{3} $.
 %
The important feature of maximal volume is that it is defined in a coordinate
invariant manner.
 To calculate the maximum volume of    $ \Sigma_{\text{total}} $, we assume the anchoring times  $ \tau_{b_{k}} $s are equal, so it is sufficient to find in an exterior chart a surface   with maximal volume and $ \mathcal{V}_{\Sigma_{\text{total}}}$ becomes $\mathcal{V}_{\Sigma_{1}}+\mathcal{V}_{\Sigma_{2}}+\mathcal{V}_{\Sigma_{3}}.   $

The extremization procedure of $ \mathcal{V}_{\Sigma_{k}} $ leads to the following complicated differential equation:
\begin{align}\label{dif}
2 \left(M+\rho ^2\right)\tau_{k} ''(\rho )+8 \rho ~ \tau_{k} '(\rho )-4 \rho  \left(M+\rho ^2\right)^2 \tau_{k}
'(\rho )^3 \text{sech}^2\left(\sqrt{M} \tau_{k} (\rho )\right)=\frac{\sqrt{M} \sinh \left(2 \sqrt{M} \tau_{k} (\rho )\right)}{\left(M+\rho ^2\right) },
\end{align}
and  its  solution  is:
\begin{equation}\label{rengerho}
	\tau_{k}(\rho )=\frac{1}{\sqrt{M}}\text{arctanh}\left( \frac{\rho \tanh\left(\sqrt{M} \tau_{b_{k}}\right) }{
		\sqrt{M+\rho ^2}}\right). 
\end{equation}
The boundary conditions derived considering the fact that the codimension one  slice should touch the boundary at the desired time, i.e.
$ \tau_{k}(\rho\to\infty )=\tau_{b_{k}} $.
Besides, for every boundary time $ \tau_{b_{k}}, $ the maximal   codimension one space-like geodesic  \eqref{rengerho} should be perpendicular to the AdS boundary.
 The first order derivative of $ \tau(\rho) $ with respect to $ \rho $ leads to:
\begin{equation}\label{roengo}
	 \tau_{k} '(\rho )=\frac{\sqrt{M} \tanh \left(\sqrt{M} \tau_{b_{k}} \right)}{\sqrt{M+\rho ^2} \left(\rho
	 	^2 \text{sech}^2\left(\sqrt{M} \tau_{b_{k}} \right)+M\right)},
\end{equation}
and the above equation vanishes at the boundary.

Differential equation \eqref{dif} is nonlinear in the  unknown function $	\tau_{k}(\rho )  $. The essential questions of existence, uniqueness and well posedness of initial and boundary value problems for nonlinear differential equations are challenging. Since our goal is to evaluate the complexity, it is necessary  to know whether  the extrema \eqref{rengerho} is maximum or not. Fortunately, in the papers \cite{Bon,Scarinci:2011np} using the notion of \textit{universal Teichmuller space}, the existence and uniqueness of maximal space-like hypersurface in $ \text{AdS}_{3} $ space-time for any number of asymptotic regions and  genus were proved. 
Considering this point, we continue the calculations with maximal hypersurface \eqref{rengerho} to find complexity of dual state.


Through the substitution of equations \eqref{rengerho} and \eqref{roengo} into    \eqref{volsigma} one can see:
\begin{equation} \label{volt11}
	\mathcal{V}_{\Sigma_{k}}(\tau_{b_{k}})=\int_{0}^{2\pi}\int_{\rho_{\text{min}}}^{\Lambda} \text{sech}(\sqrt{M} \tau_{b_{k}})
	~d\rho\hspace{.09cm} d\phi.
\end{equation}
To determine the range of integration for each part, one should find the  minimum value of $ \rho $. In general, for each of outer chart, the domain of integration is different. As already alluded to, for simplicity, all $M_{k}  $s and all boundary times $ \tau_{b_{k}} $s are equal, which  implies  a $ \mathbb{Z}_{3} $   symmetry.  Therefore,  it is sufficient to find a  $\rho_{\text{min}}^{\text{sym}}$ for one chart and the total volume becomes:\footnote{Superscript "sym" emphasizes our geometry has  $ \mathbb{Z}_{3} $ symmetry and lower bound of integral equals for each chart. }
\begin{equation}\label{ww}
	\mathcal{V}_{\Sigma_{\text{total}}}^{\text{sym}}(\tau_{b})= \frac{6\pi\Lambda}{\text{cosh}(\sqrt{M}\tau_{b} )}-\frac{3}{\text{cosh}(\sqrt{M}\tau_{b} )}\int_{0}^{2\pi} \rho_{\text{min}}^{\text{sym}}(\tau_{b},\phi )
	~ d\phi.
\end{equation}
The $ \rho_{\text{min}}^{\text{sym}} $ should be located in the overlapping region,  therefore to find its location, one must use
first relation in \eqref{rho}. Also $ \rho_{\text{min}}^{\text{sym}} $ is on the codimension one hypersurface  \eqref{rengerho}, by substituting $ \tau $ with  the equation of hypersurface, first relation in \eqref{rho} becomes:
\begin{equation}\label{overlap}
	\tilde{\rho}_{\text{min}}=-\sinh (A) \cosh \left(\sqrt{M} \phi \right) \sqrt{M+\rho ^2_{\text{min}}
	\text{sech}^2\left( \sqrt{M} \tau_{b}\right) }-\rho_{\text{min}} \cosh (A),
\end{equation}
and $ A $  can be computed from \eqref{a} as following:
\begin{equation}
	A=	\text{sech}^{-1}\left(1-\text{sech}\left(\pi  \sqrt{M}\right)\right).
\end{equation}
Since our geometry has $ \mathbb{Z}_{3} $ symmetry, one can not distinguish between $ \rho_{\text{min}} $ and $ \tilde{\rho}_{\text{min}} $. So in the left hand side of equation \eqref{overlap} we make a replacement $\tilde{\rho}_{\text{min}}\rightarrow \rho_{\text{min}} $ and solve it to find $ \rho_{\text{min}}^{\text{sym}} $  as:
\begin{equation}\label{rmin}
\rho_{\text{min}}^{\text{sym}}(\tau_{b},\phi )=\frac{\sqrt{M} \sinh (A) \cosh \left(\sqrt{M} \phi \right)}{\sqrt{(\cosh
		(A)+1)^2-\sinh ^2(A) \text{sech}^2(\sqrt{M} \tau_{b}) \cosh ^2\left(\sqrt{M} \phi \right)}}.
\end{equation}
From the above relation it is obvious that $ \rho_{\text{min}}^{\text{sym}} $ depends on $ \phi $ coordinate and boundary time $ \tau_{b} $, but for some range of $ \phi $ and $ \tau_{b} $,  $ \rho_{\text{min}}^{\text{sym}} $ is imaginary. Thus, for any  $ \phi\in[0,2\pi) $,  the range of  $ \tau_{b} $ is:
\begin{equation}\label{rengt}
\frac{1}{\sqrt{M}}\cosh ^{-1}\left(\frac{\cosh \left(2 \pi  \sqrt{M}\right)}{\sqrt{2 \cosh
		\left(\pi  \sqrt{M}\right)-1}}\right)<\mid\tau_{b}\mid.
\end{equation}
After substituting \eqref{rmin} in to \eqref{ww} we have to perform the integral over $\phi  $. Using
 \begin{equation}
\int\frac{1}{\sqrt{A-B x^2}} dx=\frac{1}{\sqrt{B}}\tan ^{-1}\left(\frac{\sqrt{B} x}{\sqrt{A-B x^2}}\right),
 \end{equation}
we learn, after setting $x= \sinh \left(\sqrt{M} \phi \right)  $ that: 
\begin{equation}\label{ew}
	\mathcal{V}_{\Sigma_{\text{total}}}^{\text{sym}}(\tau_{b})= \frac{6\pi\Lambda}{\text{cosh}(\sqrt{M}\tau_{b} )}-
 3\tan ^{-1}\left(\mathcal{A}\right),
\end{equation}
where
\begin{dmath}
\mathcal{A}=\frac{\sqrt{2} \sinh (A) \sinh \left(2 \pi 
	\sqrt{M}\right)}{\sqrt{(\cosh (A)+1) \left((\cosh (A)+1) \cosh \left(2
		\sqrt{M} \tau_{b} \right)-\left((\cosh (A)-1) \cosh \left(4 \pi 
		\sqrt{M}\right)\right)+2\right)}}.
\end{dmath}
As  noted in section \ref{2}, the space-time regions outside the horizons of the Lorentzian wormholes geometry  is BTZ black hole, and because we wish to calculate the complexity of the state on the boundary of  this space-time it is convenient to rewrite   \eqref{ew}  in coordinate of BTZ black hole i.e., $ (t_{b},r_{\infty}) $.
Therefore, in the limit $ \rho\rightarrow\Lambda $ from relations \eqref{bt1} and \eqref{bt}, one can conclude that 	
$   \tau_{b}\simeq t_{b} $, and $ r\simeq\frac{\Lambda}{\cosh(\sqrt{M}t_{b})} $.
Due to the fact that  the fate of state which is located on the boundary $ (r=r_{\infty})  $ is important, $ \Lambda $ should be chosen very large such that at late times we also have $ \frac{\Lambda}{\cosh(\sqrt{M}t_{b})}\gg1 $. So the complexity of state \eqref{tripartite} becomes:
\begin{dmath}\label{cew}
	\mathcal{C}_{\text{V}}\left(	|\Psi \rangle \right)=6\pi r_{\infty}-
 3\tan ^{-1}\left(\mathcal{A}\right).
\end{dmath}

 Also the complexity of AdS$ _{3} $ is
\begin{equation}
	\mathcal{C}^{\text{AdS}}_{\text{V}}=2\pi r_{\infty}-2\pi+\mathcal{O}(1/r_{\infty}).
\end{equation}
Hence, the complexity of formation for three entangled black holes  is finite \cite{Chapman:2016hwi},\footnote{In \cite{Fu:2018kcp} the authors use hyperbolic slicing which means that they consider the metric as $ ds^2 = -dt^2 +\cos^2t~ d^2\Sigma_2 $ where $ \Sigma_2  = H_{2}/\Gamma$ and then compute the complexity of formation via CV and CA proposals. But in our case, the geometry is given by $ H_{3}/\Gamma $. Accordingly, their geometries are actually non-handlebodies but our case is handlebody.  Furthermore, they have a global time coordinate but in our case, it does not exist.}
\begin{equation}\label{fin}
	\Delta\mathcal{C}_{\text{V}}=	\mathcal{C}_{\text{V}}\left(	|\Psi \rangle \right)-3\mathcal{C}^{\text{AdS}}_{\text{V}}=6\pi+\text{finit terms}.
\end{equation}
If   $ M $  is taken to be large, finite terms in \eqref{fin} will have the simple form
\begin{equation}\label{cform}
	\Delta\mathcal{C}_{\text{V}}\simeq	6 \pi -3 \tan ^{-1}\left(\frac{e^{\frac{3 \pi  \sqrt{M}}{2}}}{\sqrt{2 \cosh
			\left(2 \sqrt{M} t_{b}\right)-e^{3 \pi  \sqrt{M}}}}\right),
\end{equation}
and in  this limit,  \eqref{rengt}  becomes  $\mid\tau_{b}\mid>\frac{3 \pi }{2}$.  Expanding  \eqref{cform} around $ t_{b}\sim\frac{3 \pi }{2} $  leads to  
\begin{equation}
3 e^{3 \pi  \sqrt{M}} \sqrt{M} t_{b}-\frac{9}{2} \pi  e^{3 \pi  \sqrt{M}}
\sqrt{M}-3 \tan ^{-1}\left(e^{3 \pi  \sqrt{M}}\right)+6 \pi+\mathcal{O}\left(\left(t_{b}-\frac{3 \pi }{2}\right)^2\right),
\end{equation}
 and it can be seen that the complexity grows linearly at early times, but at longer times this growth is not linear.
 
\begin{figure}[h]
	\centering
	\begin{overpic}
		[width=0.9\textwidth,tics=11]{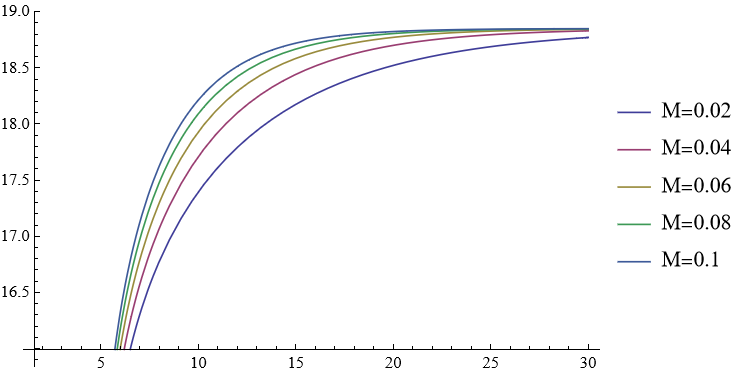}
		\put (0,52) {$\displaystyle\Delta\mathcal{C}_{V}$}
		\put (82,2) {$\displaystyle t_{b}$}	
	\end{overpic}
	\caption{Plots of complexity of formation as a function of $t_{b}  $, for some values of
		 $ M $. Complexity of formation  grows non linearly  and then saturates at late times $ (\frac{1}{\sqrt{M}}\ll t_{b}) $. According to \eqref{rengt}, 	in the limit $ M\ll 1 $, $ t_{b} $ should be greater than $  \sqrt{3} \pi -\frac{\sqrt{3}}{8}  \pi
		 ^3 M$, so the curves related to smaller $ M $s begin at a farther  distance from the origin. }	
	\label{volppp}
\end{figure}

  At late times (i.e.  $\frac{1}{\sqrt{M}}\ll t_{b} $) \eqref{cform} becomes:
 \begin{equation}\label{lform}
 	\Delta\mathcal{C}_{\text{V}}\simeq	6 \pi -3 \tan ^{-1}\left(\frac{e^{\frac{3 \pi  \sqrt{M}}{2}}}{e^{\sqrt{M}t_{b}}}\right),
 \end{equation}
and approaches to a constant value $ 6\pi $. 
 Figure \ref{volppp} shows some plots of $ 	\Delta\mathcal{C}_{\text{V}} $ as a function of anchoring time $t_{b}  $ for some different values of $ M $.
One can see that the complexity grows non-linearly  and then saturates  at late times. In other words, the complexity of this geometry indicates three important features. First, it grows over time secondly, this growth is nonlinear, and thirdly, it saturates at late times.
According to the second law of complexity \cite{Brown:2017jil}, the first feature is consistent with the common behavior of complexity. While the second and third features show deviation from the known behaviors.
For example, both  Einstein-Rosen bridge in two-sided AdS eternal black hole and bridge-to-nowhere in one-sided AdS black hole behave in the same manner and increase linearly. In classical gravity, Einstein-Rosen bridge and bridge-to-nowhere grow forever, so there is no upper bound on holographic complexity. The problem is that  there are only a finite number of orthogonal
states and after an exponential time we finish orthogonal states and quantum complexity reaches equilibrium. It is said that classical general relatively breaks down at time $\sim\exp(S)$ so this inconsistency is resolved\cite{Susskind:2018pmk}.
Interestingly, Lorentzian wormhole  geometry shows the saturation of complexity even in classical gravity. Therefore, the third feature can be considered a virtue, while the second feature is not compatible with Lloyd's bound or linear growth.

It is noteworthy  to mention that, since  the complexity increases, one can conclude the horizons of multi-black holes are transparent and when it saturates they become multi-gray holes. Also, notice that complexity of state \eqref{cew} is an even function of time so the complexity increases in to the past $ (\tau_{b}<\frac{-1}{\sqrt{M}}\cosh ^{-1}\left(\frac{\cosh \left(2 \pi  \sqrt{M}\right)}{\sqrt{2 \cosh
			\left(\pi  \sqrt{M}\right)-1}}\right)) $ hence, we have multi-white holes which their horizons are opaque \cite{Susskind:2015toa}.

\vspace{.5cm}
\newpage
\section{Conclusion} \label{4}

 Multi-black holes geometries  which can be constructed out of pieces of AdS spacetime are non-perturbative three-dimensional vacuum solutions of
 Einstein-Hilbert theory.
  These geometries are generalization of BTZ black hole which have asymptotically AdS exterior regions that join in
  one interior region. The geometry of each exterior region has a black hole horizon.
  
Understanding the behavior of the volume of the semi-classical black hole behind the horizon at very late times  is an important goal in quantum gravity. Therefore, in this paper, an attempt has been made to investigate the holographic complexity for the simplest generalization of BTZ black hole i.e. $ 2+1 $ dimensional Lorentzian wormholes geometry with three asymptotic AdS boundaries. The corresponding state is similar to $ \text{GHZ} $ state, represents tripartite entanglement.
 In our calculation, it was assumed all horizons are equal and the maximal Cauchy slice anchored to the asymptotic boundaries at equal times.
 
 It was seen that the complexity of the dual state of this space-time  grows with time nonlinearly and then saturates at  late times ($ \frac{1}{\sqrt{M}}\ll t_{b} $), so  Lloyd’s
 bound is not satisfied.
 This result does not show inconsistency with the main features of complexity, i.e. its tendency to increase (second law of complexity)  and its saturation.
 In BTZ geometry the volume of the wormhole grows forever, and  it is not compatible with quantum complexity behavior. Because there is a limited number of orthogonal states, the number being of order $ \exp(S) $, and complexity should saturate when $ t\sim\exp(S) $, we say that the description of BTZ geometry as a solution for classical geometry must break down at  $ t\sim\exp(S) $ \cite{Susskind:2018pmk}. Based on this 
 note, it seems one can  see the saturation of complexity in classical gravity, as well.

To make the calculation tractable, we assumed that the $ \mathbb{Z}_{3} $ symmetry  exists, but it is expected that breaking this symmetry by choosing different anchoring times does not affect the overall behavior of complexity. Similar to    the two-sided eternal BTZ black hole case \eqref{thermo}, choosing different  $ t_L $ and $ t_R $, or even varying $ t_R $ with fixed  $ t_L $    does not significantly change the linear growth. It is worth noting that in the two-sided eternal BTZ black hole there is  boost symmetry and when this symmetry is broken ($ t_R\neq-t_L $ )  the complexity begins to grow while multi boundary wormholes have no globally defined Killing vector fields.
 However, it would be interesting to find complexity for different $ M_k $s and $ t_{b_{k}} $s.


\subsubsection*{Acknowledgments}
Special thanks to Ali Naseh for the discussions and encouragement. The author would like to extend his appreciation to Mohsen Alishahiha, Mostafa Ghasemi, Ghadir Jafari, Kirill Krasnov, Amin Talebi, and Balt C.van Rees  for constructive discussions
on related subjects.

\end{document}